\documentclass[preprint]{elsarticle}

\usepackage{graphicx}
\usepackage{amsmath,amssymb,epsfig}
\usepackage{multirow}
\usepackage{tikz}

\newcommand{\rd}{\mathrm{d}}

\newcommand{\td}[2]{\frac{\rd #1}{\rd #2}}

\newcommand{\beq}{\begin{equation}}
\newcommand{\eeq}{\end{equation}}

\begin{document}
\begin{frontmatter}



\title{Hospitalization dynamics during the first COVID-19 pandemic wave: SIR modelling compared to Belgium, France, Italy, Switzerland and New York City data}


\author{Gregory Kozyreff}
\ead{gkozyref@ulb.ac.be}
\address{Optique Nonlin\'eaire Th\'eorique, Universit\'e libre de Bruxelles (U.L.B.), CP 231, Belgium}
\date{\today}

\begin{abstract}
Using the classical Susceptible-Infected-Recovered epidemiological model, an analytical formula is derived for the number of beds occupied by Covid-19 patients. The analytical curve is fitted to data in Belgium, France, New York City and Switzerland, with a correlation coefficient exceeding 98.8\%, suggesting that finer models are unnecessary with such macroscopic data. The fitting is used to extract estimates of the doubling time in the ascending phase of the epidemic, the mean recovery time and, for those who require medical intervention, the mean hospitalization time. Large variations can be observed among different outbreaks.
\end{abstract}


\end{frontmatter}


\section{Introduction} 

As the COVID-19 pandemic continues to develop in various parts of the world, the scientific community at large is producing a massive effort to gather as many information as possible on the nature of virus and its ability to spread. At the epidemiological level, some parameters are of course particularly desirable to ascertain such as the rate of infection, $\beta$, within a given population and the average time of  spontaneous recovery, $t_R$. But perhaps even more important from a crisis management point of view is to be able to predict how many beds in hospital are needed and how long they will be occupied. Some estimates are of course already available \cite{Zhou-2020,Li-2020,Chowdhury-2020,Chen-2020,Wu-2020,Richardson-2020}, however they usually rest on the analysis of rather small cohorts. This paper is an attempt to extract such an information from the data made available in several similar but different contexts: Belgium, France, Italy, Switzerland, and New York City (NYC). Note that NYC contains a well-delimited population whose size is comparable to those of Switzerland or Belgium and, hence, can be treated on an equal footing as a small country from an epidemiological point of view.  It will be shown that an  analytical curve derived from the simplest possible SIR compartment model can be made to fit remarkably well with the data, with a correlation coefficient ranging from 0.988 in New York City to beyond 0.997 elsewhere. With such good fits, the obtained curves can safely be extrapolated to forecast bed occupation several weeks in advance. This calls into question the necessity to resort to more complex modelling, involving more parameters than the SIR model, to confront macroscopic data in the absence of more detailed information~\cite{Giordano-2020,Roda-2020,Gatto-2020}. Note that SIR modelling has been applied to several outbreaks of COVID-19 in various parts of the world~\cite{Calafiore-2020,Bagal-2020,Simha-2020,Bastos-2020}. However, at the time of their publication, the local outbreak was still in the ascending phase and the wave was therefore not complete, which limited the accuracy of data fitting.  The study also shows that a great disparity of epidemiological parameters can  exist between different countries, despite their similarities. This reflects both the policies put in places to mitigate local epidemics and also, perhaps, underlines differences between health systems.

Besides the managerial motivation invoked above, the focus in this study is on the hospitalization dynamics for two reasons. Firstly, since the beginning of the pandemic, a large uncertainty has been surrounding the number of cases, as the ability and protocols to test patients varies from one country to the next. Estimates of the number of infected people, as well as when the epidemics started in a given region appear poorly reliable. By contrast, bed occupation numbers are easier to monitor. Next, the processes leading to the number of occupied beds are multiple and additive, leading to  a much smoother data than, for instance, the daily numbers of admitted and discharged patients. Hence, curve fitting is expected to yield more reliable information with hospitalization data. Except for NYC, where the data is lacking, we will both exploit general bed occupancy and the number of patients in Intensive Care Units (ICU).

\section{Mathematical modelling}
The simplest of all epidemiological models is the SIR model, which separates a given population into a set of susceptible (S), infected (I), and recovered  (R) individuals. These populations evolve in time according to
\begin{align}
\td St&=-\beta SI/N,\\
\td It&=\beta SI/N- I/t_R,\\
\td Rt&= I/t_R,
\end{align}
where $S(t)+I(t)+R(t)=N$, the size of the population, $\beta$ is the infection rate and $t_R$ is the spontaneous recovery time. In the majority of countries where confinement measures have been taken, the growing exponential phase of the local outbreak was stopped well before a sizeable fraction of the population was infected. Hence, thanks to public intervention, $I(t), R(t)\ll N$ at all time, even if they could reach considerable values. Therefore, one has $S(t)\approx N$ and the equation for $I(t)$ becomes, with good approximation,
\beq
\td It = \left(\beta-1/t_R\right)I.
\eeq
The effect of confinement and social distancing is to reduce the coefficient $\beta$, so that this parameter is a function of time. For simplicity, we assume that there is a well defined date at which $\beta$ switches from a large value $\beta_0$ to a smaller one, $\beta_1$. This, of course, is an approximation, but it appears acceptable since there has been, in most of the setting considered in this study, a well defined date where the local authority has declared some form of lockdown~\cite{Flaxman-2020}. Taking, for each outbreak, $t=0$ as the time when lockdown started, we thus have
\beq
I(t)= I_0\times\left\{\begin{matrix}  e^{ct}, &  t<0,\\   e^{-\Gamma t}, & t>0.\end{matrix}\right. 
\label{eq:I(t)}
\eeq
where $c$ and $\Gamma$ are given by
\begin{align}
c&= \beta_0-1/t_R, & \Gamma&= 1/t_R-\beta_1
\end{align}
and are, respectively, the initial growth rate and the late-time decay rate.   Equivalent to $c$, and more convenient to discuss, is the doubling time $t_d=\ln(2)/c$ during the initial phase of the local outbreak. 

In Eq.~(\ref{eq:I(t)}), $I_0$ is the value of $I(t)$ at $t=0$, a number difficult to determine with accuracy. Note that all we can learn from the data of $I(t)$ is $c$ and $\Gamma$, which is not quite enough to know $\beta_0$, $\beta_1$ and $t_R$. Hopefully, $\beta_1$ is close to zero, but in all probability it isn't. Hence, $t_R<\Gamma^{-1}$. However, one may hypothesize that the populations involved with the pandemic in Belgium, France, Switzerland, Italy and New York City all have similar response to the virus, so that they share the same value $t_R$. Hence, the smallest of the values of $\Gamma^{-1}$ extracted from these five epidemic events may count as the best estimation of the upper bound on $t_R$. 

Knowing $I(t)$, the evolution of the number of hospitalized patients $P(t)$ is straightforward to model. It obeys the equation
\beq
\td {P(t)}t=\alpha \Gamma I(t-\tau) -P(t)/t_H,
\label{eqP}
\eeq
which expresses, simply, that the number of hospitalisations increases at a rate proportional to the number of infected people and that, once admitted into hospital, the mean time of stay is $t_H$. Above, $\alpha$ is the probability, if infected, to be hospitalised. In this last Eq., $I(t)$ appears with a delay $\tau$. This delay accounts, for the most part, for the average time elapsed between being infected and requiring to be hospitalized; additionally, one may conjecture  that social event such as mass gatherings may have further delayed the response to the measures, leading to a larger value of $\tau$.

Combining Eqs.~(\ref{eq:I(t)}) and (\ref{eqP}), one derives
\begin{align}
P(t)&= p \, e^{c(t-\tau)}\left(1-e^{-(ct_H+1)(t-t_0)/t_H}\right), &t<\tau, \label{P:1}\\
&= p\left[\left(1-e^{-(ct_H+1)(\tau-t_0)/t_H}\right)e^{-(t-\tau)/t_H} +\frac{ct_H+1}{\Gamma t_H-1}\left(e^{-(t-\tau)/t_H}-e^{-\Gamma (t-\tau)}\right)\right], &t>\tau,
\label{P:2}
\end{align}
where $t_0$ is the time of the first hospital admission and where $p=\alpha I_0 e^{c\tau}\Gamma t_H/(1+ct_H)$. Finding $p$, it would be particularly interesting to deduce $\alpha$. Unfortunately, this requires the knowledge of $I_0$, which we don't have.

In the same way as for $P(t)$, one may derive an evolution model for the number of occupied beds in Intensive Care Unit (ICU), $P_{ICU}(t)$. The simplest way is to write
\beq
\td{P_{ICU}(t)}t=\alpha_{ICU} \Gamma I(t-\tau') -P_{ICU}(t)/t_{ICU}.
\label{eq:ICU}
\eeq
The above equation neglects intermediate stages between being infected and integrating the ICU. Accordingly, the evolution of $P_{ICU}(t)$ is given by the same expressions as in Eqs.~(\ref{P:1}) and (\ref{P:2}) but with the substitutions $t_H\to t_{ICU}$, $\tau\to\tau'$, and  $p\to p_{ICU}$.

One may argue that Eqs.~(\ref{eq:I(t)}) to (\ref{eq:ICU}) are oversimplified in that the model neglects an intermediate population $E(t)$ of exposed, not-yet contagious individuals, and that the population $P_{ICU}$ should rather be coupled to the larger set $P(t)$ rather than $I(t)$ as in~\cite{Chowdhury-2020}.  In the same vein, we have not separated the population in age categories, even though this would be highly relevant~\cite{Onder-2020}.
 However, the attitude in the present paper is to invoke the simplest possible model in order to exploit simple explicit formulas like Eqs.~(\ref{P:1}) and (\ref{P:2}). As we will see, this yields excellent fit to the data.

\section{Data and Fit}

Hospitalization data was gathered  for
\begin{enumerate}
\item Belgium, during the period extending from 2020-03-15 to 2020-05-28 \cite{Sciensano-2020}. Official lockdown was imposed on 2020-03-18. The first patient was hospitalized on 2020-02-04.
\item France, during the period extending from 2020-03-18 to 2020-05-28 \cite{France-2020}. Official lockdown was imposed on 2020-03-17. The first patient was hospitalized on 2020-01-24.
\item Italy, during the period extending from 2020-02-24 to 2020-05-28 \cite{Italy-2020}. Official lockdown was imposed on 2020-03-9. Estimation of the first hospitalization is 2020-02-07.
\item Switzerland, during the period extending from 2020-02-25 to 2020-05-28  \cite{Switzerland-2020}. Containment measures were taken as of 2020-03-20. First hospitalization was on 2020-02-26.
\item New York City, during the period extending from 2020-02-29 to 2020-05-23  \cite{NYC-2020}. Stay-at-home order was enforced on  2020-03-22. Estimation of the first hospitalization is 2020-02-29.
\end{enumerate}

The data sets were analyzed with Mathematica 8.0 using the functions `FindFit' and `NonlinearModelFit'. Both commands allow to obtain parameter set by least-square regression and the latter yields 95\% confidence intervals (CI). One notes that the parameters $\Gamma^{-1}$ and $t_H$ (or $t_{ICU}$) appear in separate but very similar mathematical terms in Eq.~(\ref{P:2}), making these two parameters strongly correlated and rendering their determination ambiguous. For example, with the french data, the pairs $(\Gamma^{-1},t_H)\approx(12,42)$ and  $(\Gamma^{-1},t_H)\approx(42,12)$ can be made to yield almost equally good fit. In order to remove the ambiguity and narrow down 95\% CI, one should fix one of these two parameters. To this end, a preliminary round of parameter fitting was carried out for each geographical region in order to determine sensible values for $\Gamma^{-1}$. In this first round, $t_d$, the doubling time in the initial phase, was varied between 3 and 7 and imposed prior to fitting, while $\Gamma^{-1}$ and $t_H$ were free to vary. The best-fitting value of $\Gamma^{-1}$ was thus monitored as a function of $t_d$. For the Belgian data $\Gamma^{-1}$ was always close to 16 days. For Italy $\Gamma^{-1}$ was found to lie between 15 and 21 days, whereas NYC and Switzerland gave lower values, around 9 days. From this initial investigation, one makes the following informed guesses: $\Gamma^{-1}(\text{Belgium})=16$, $\Gamma^{-1}(\text{Italy})=20$, $\Gamma^{-1}(\text{NYC, Switzerland})=9$. Finally, for France, one assumes the same value as in Belgium: $\Gamma^{-1}(\text{France})=16$. Adopting these values, a new round of parameter fitting was carried out in which $\Gamma^{-1}$ was fixed and all other parameters were free to vary. What is output below as 95\% IC in the following should of  course be regarded, at best, as conditional probabilities. They are mere indicators of uncertainty. Note that  a significantly longer characteristic time $\Gamma^{-1}(Italy)$ can be deduced from the study of Giordano {\it et al.}~\cite{Giordano-2020}, but their estimate rests on data covering a much shorter period of time, in which the outbreak was still in the ascending phase,  whereas here the full epidemic wave is taken into account.

Curve fitting was done separately with ICU data.

Given the set of data points $(t_i,y_i)$ for a given outbreak, with mean value $\bar y$, the correlation coefficient was computed as
\beq
\mathcal C= \frac{\sum_i \left(P(t_i)-\bar y\right)\left(y_i-\bar y\right)}
{\sqrt{\sum_i \left(P(t_i)-\bar y\right)^2}\sqrt{\sum_i \left(y_i-\bar y\right)^2}}.
\eeq

\section{Results and discussion}

\begin{table}
\caption{Fitting values (95\%CI) of doubling times $t_d$ in the growing phase, effective lock-down times ($\tau, \tau'$), hospitalization times ($t_H, t_{ICU}$) and coefficient $p$. The characteristic time $\Gamma^{-1}$ of the exponential decreasing phase of the outbreak was set to a fixed value for each country or city. Times are expressed in days.}\label{tab:fit}
\begin{tabular}{llllll}
\hline\noalign{\smallskip}
                     &$t_d$& $p, p_{ICU}$           & $\tau, \tau'$  & $ t_{H}, t_{ICU}$  & $\Gamma^{-1}$  \\
\noalign{\smallskip}\hline\noalign{\smallskip}
Belgium         & 3.5 - 4.1  & 2920 - 3210 &  7.5 - 8.3 & 16.1 - 17.0 & 16 \\
Belgium (ICU) & 3.5 - 4.1    & 671 - 739    &  8.0 - 8.8 & 14.8 - 15.7 & 16   \\
France           & 4.0 - 4.5  & 13970 - 15250 & 9.1 - 9.8 & {\bf 34.3 - 35.7} & 16  \\
France (ICU)   & 4.4 - 5.0  & 4010 - 4330 &  10.4 - 11.3& 14.9 - 15.7 & 16  \\
Italy               & 5.4 - 6.2  & 19110 - 20630  &  10.9 - 12.0 & 19.1 - 20.4 & 20 \\
Italy (ICU)       & 5.8 - 6.7    & 2701 - 2907 &  11.2 - 12.3 & 11.5 - 12.7 & 20   \\
Switzerland	 & 4.2 - 4.9  & 2780 - 2997 & 3.1 - 4.0  &  20.5 - 21.8 & 9 \\ 
Switzerland (ICU) &4.9 - 5.6 & 549 - 582 &  6.6 - 7.4  &  15.1 - 16.2 & 9  \\
NYC	          & 3.4 - 4.7  & 1006 - 1176 &  0.1 - 1.7  &  {\bf 11.6 - 13.6} & 9  \\
\noalign{\smallskip}\hline
\end{tabular}
\end{table}

\begin{figure}
\includegraphics[width=7cm]{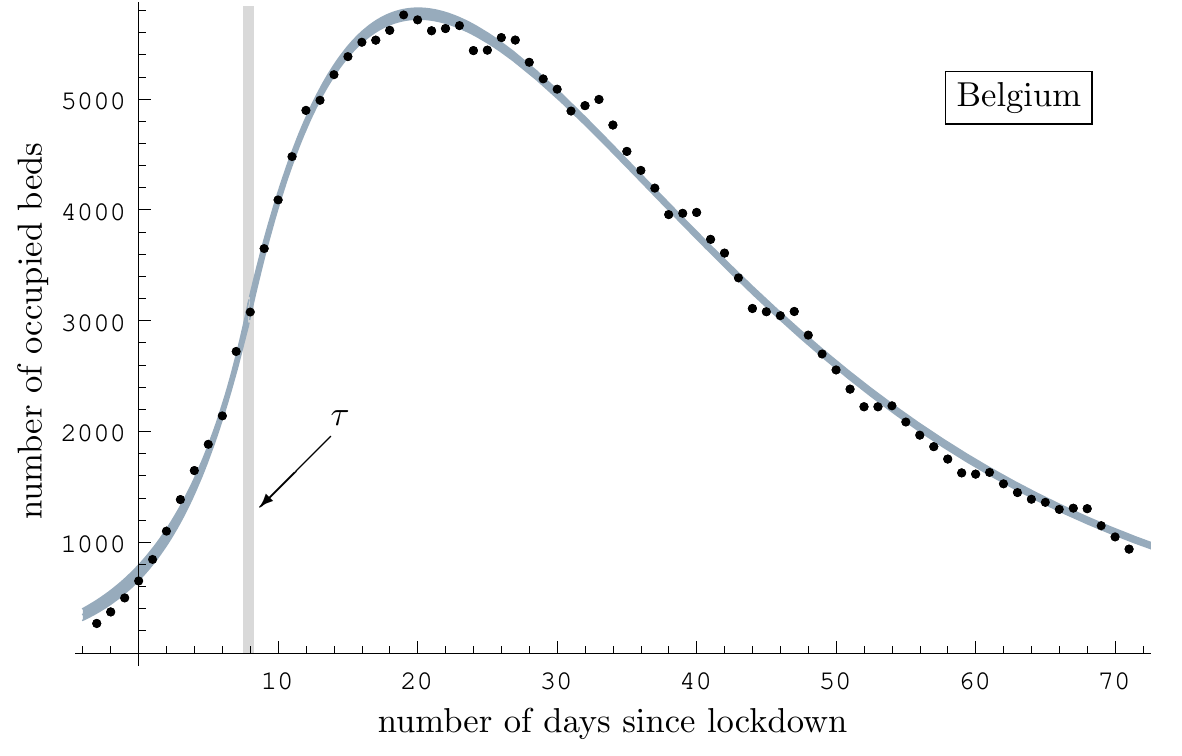}
\includegraphics[width=7cm]{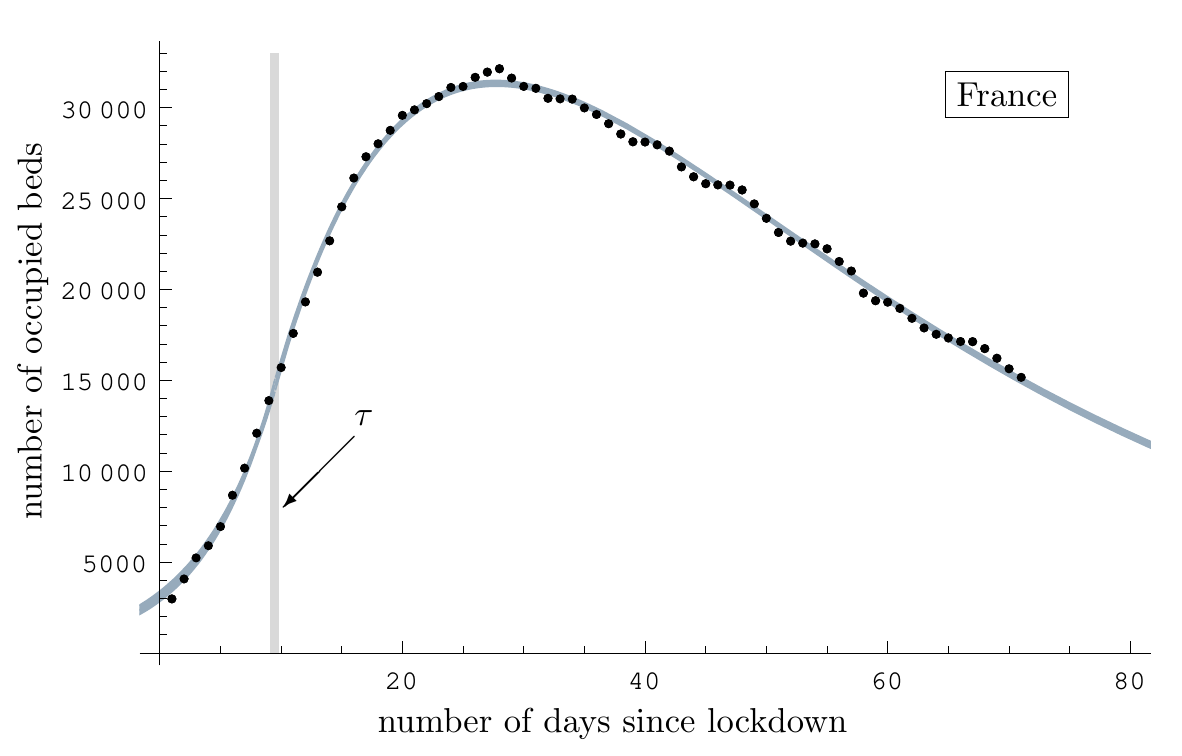}
\includegraphics[width=7cm]{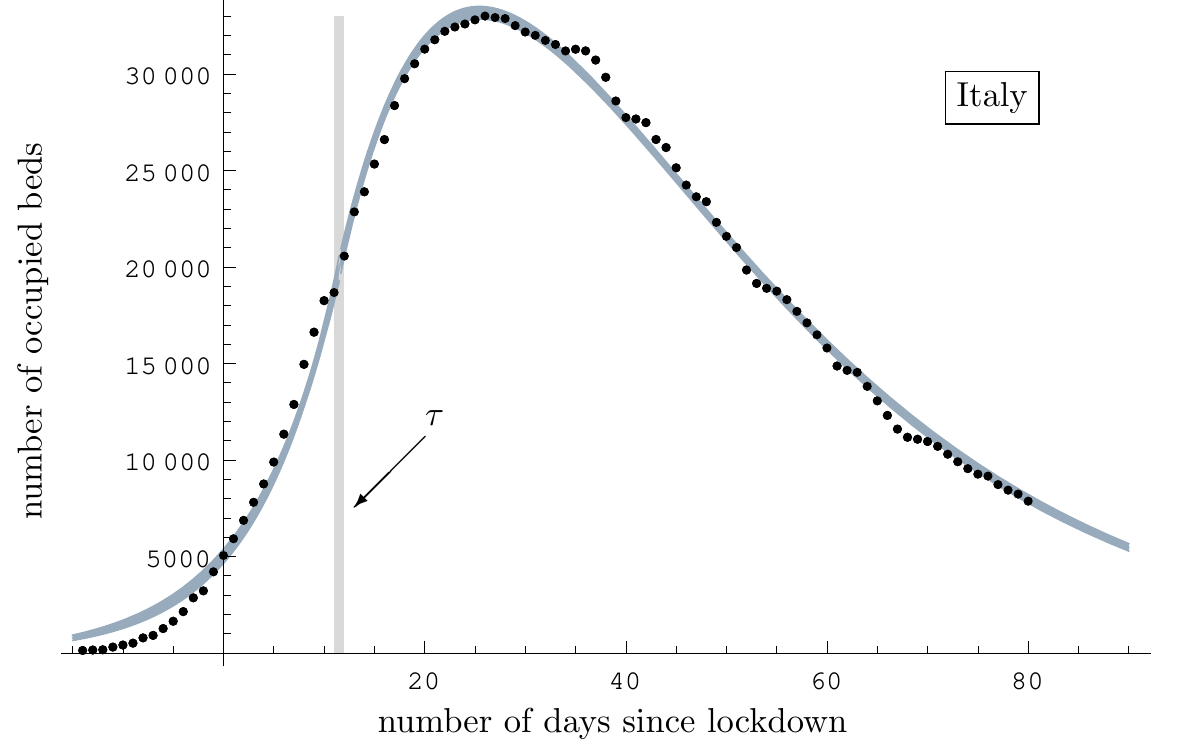}
\includegraphics[width=7cm]{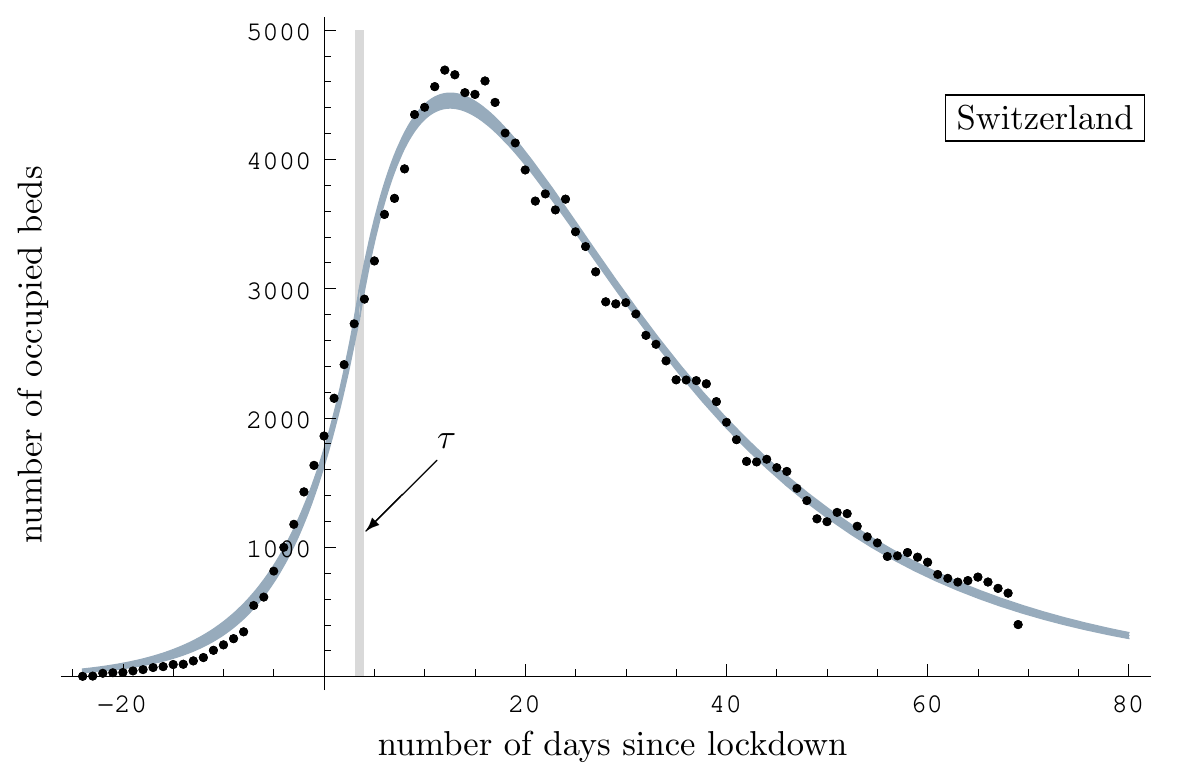}
\includegraphics[width=7cm]{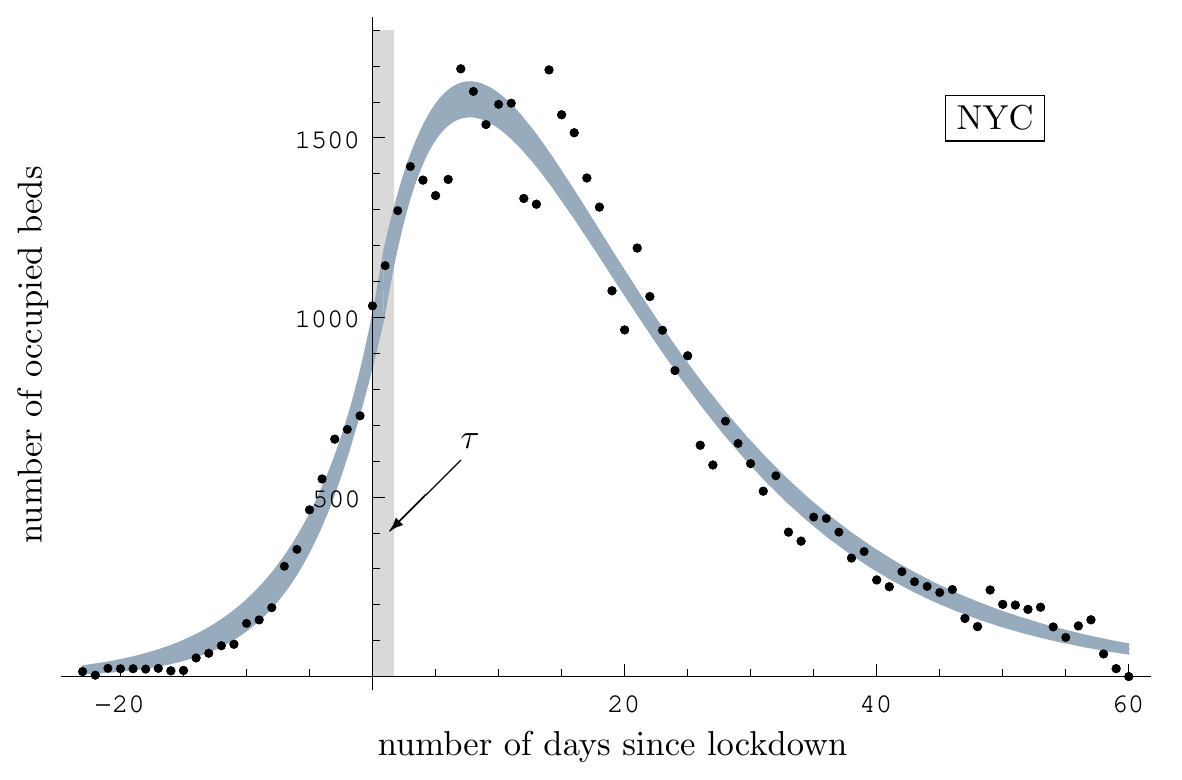}
\caption{Number of occupied beds as a function of time in general hospitalization. Dots: data points communicated by official agencies. Thick curves: 95\% confidence band using the analytical model. Gray lines indicate the value of $\tau$.}\label{fig:compare:all}
\end{figure}

\begin{figure}
\includegraphics[width=7cm]{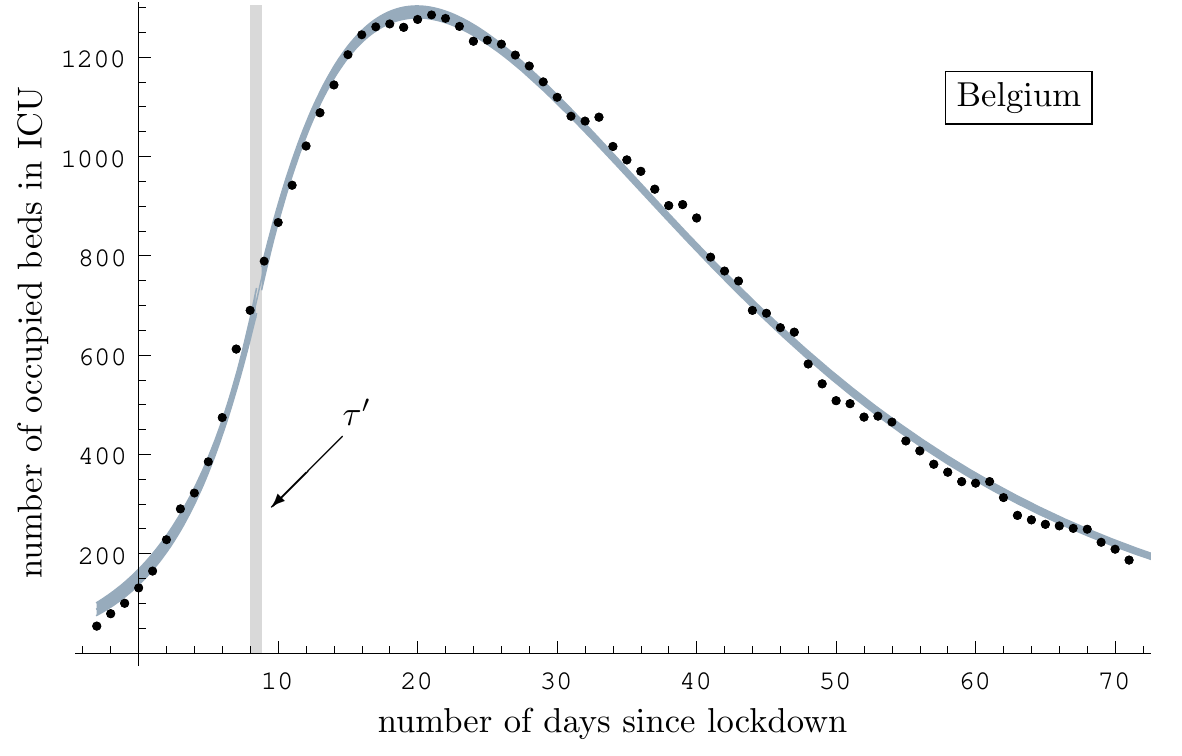}
\includegraphics[width=7cm]{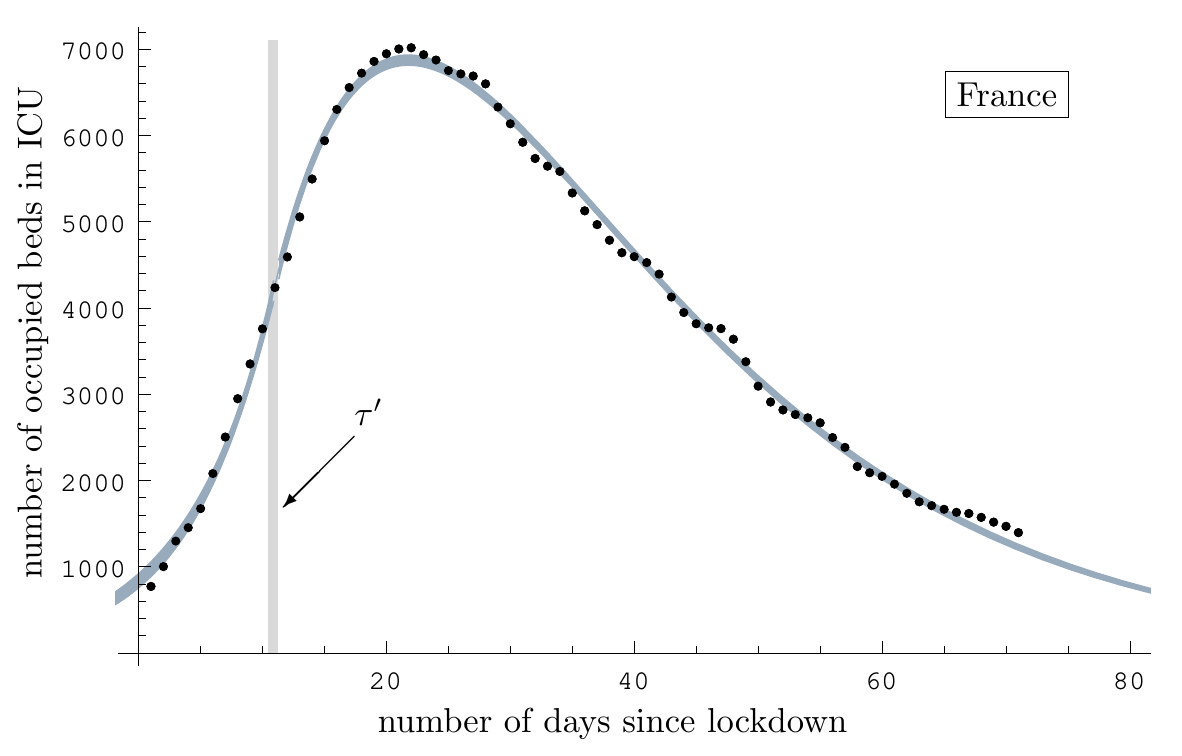}
\includegraphics[width=7cm]{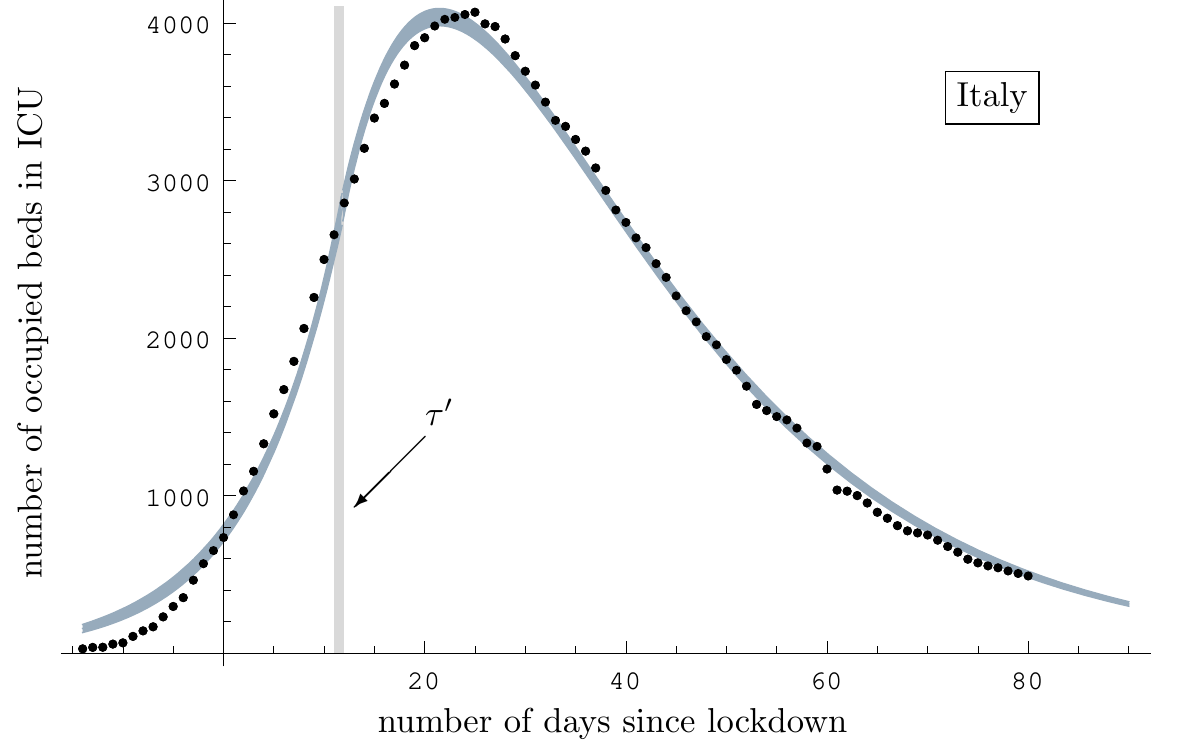}
\includegraphics[width=7cm]{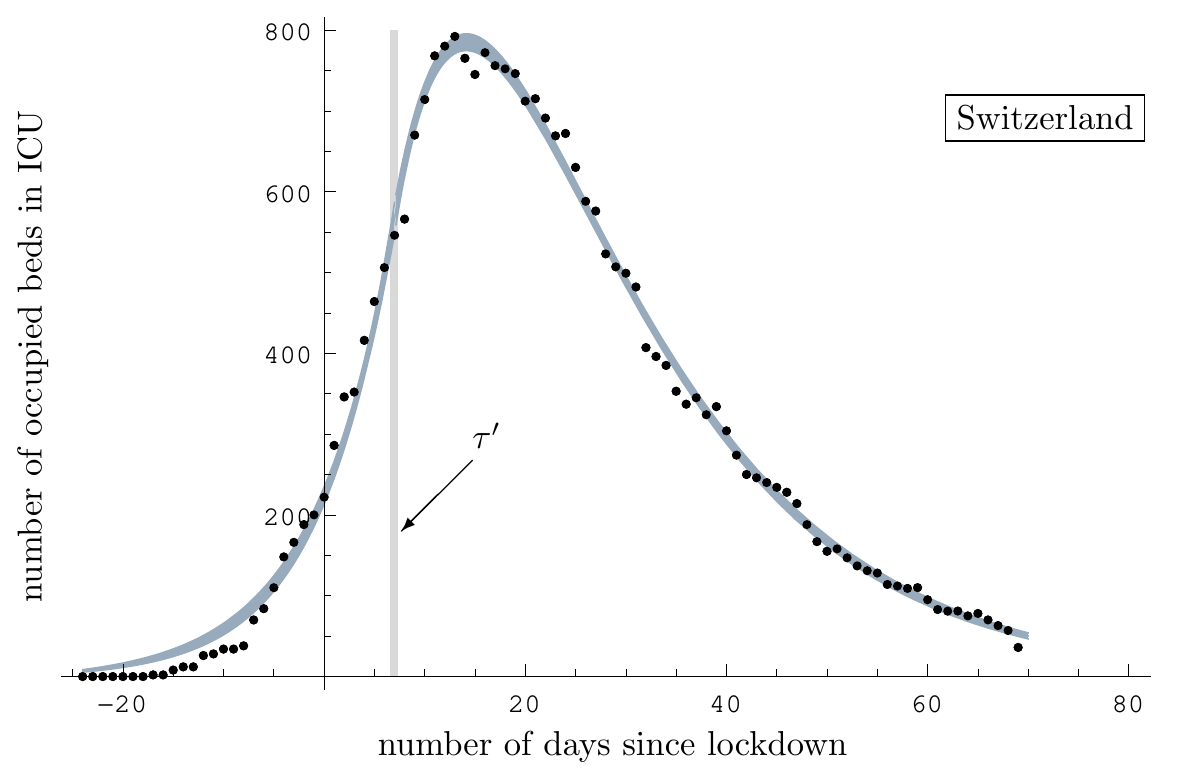}
\caption{Number of occupied beds as a function of time in ICU. Dots: data points communicated by official agencies. Thick curves: 95\% confidence band using the analytical model. Gray lines indicate the value of $\tau$.}\label{fig:compare:all2}
\end{figure}

The comparisons of the analytical curves with the official data is shown in Figs.~\ref{fig:compare:all} and~\ref{fig:compare:all2}. In all cases, a close fit is obtained with the analytical formula, with $\mathcal C$ almost equal to 1.  A close inspection of the curves shows that the growth phase is not purely exponential, meaning that $\beta$ is not simply a constant $\beta_0$ during that phase. This was anticipated. The ranges of values for the various parameters are summarized in Table~\ref{tab:fit}. Notable similarities, but also differences, can be seen from one country/city to another. Below, we highlight some of them.

Belgium and NYC have been exposed to the most rapidly growing outbreaks with doubling time under 4 days; this  is consistent with their high population densities. $t_d$ was between 4 and 5 days in France and Switzerland, and between 5 and 6 days in Italy.

Italy, France, and Belgium imposed very similar lockdown measures with very similar restrictions. Their ranges of values for $\tau$ are also similar in their lower bound. However, one clearly sees that the response to the lockdown measures was delayed by a longer time in Italy and in France than in Belgium. The larger value of $\tau$ in Italy than in France may be due to the fact that lockdown was imposed in two steps: on the 8th of March in the northern region and the next day to the rest of the country. In the case of France, one recalls that the first round of local elections took place two days before the lockdown all over the country. Despite precautions, this may have given a last-minute boost to the outbreak, accounting for the larger values of $\tau$. Ranges of values obtained with ICU data are systematically shifted towards longer times, indicating an extra delay between the development of severe symptoms and the further degradation of health condition necessitating ICU care.

The Belgian CI for $\tau$  suggests a time from exposition to severe symptoms of $7.9\pm0.4$ days. This appears consistent with previous reports. In a retrospective study of clinical progression in patients of a single helath center in Shangai (Jan 20 to Feb 6 2020) Chen \textit{et. al} report a time from onset of symptoms to hospital of 4 (2-7) days~\cite{Chen-2020}, while an incubation time of 5.1 days (4.5-5.8) has been reported~\cite{Lauer-2020,Boldog-2020}. 

In all cases where ICU data were available, $t_{ICU}$ is significantly less than $t_H$: by approximately 1 day in Belgium, 20 days in France, 8 days in Italy and 5 days in Switzerland. Combining Belgium, Italy and Switzerland, $t_H$ lies in the range 16-22 days. The study in Shanghai (China) reports a similar number for discharge time: 16 days (12-20)~\cite{Chen-2020}. With an hospitalization time of $35\pm0.7$ days, France  appears as an outlier. The opposite is seen with NYC data, which show strikingly shorter average stay at hospital that in the European countries considered. This suggests two opposite management policies: the former (France as an extreme case) where patients are kept until complete recovery, the latter (NYC) in which patient are discharged as soon as possible in order to make way for new admissions. The present fit estimates $t_H(\text{NYC})= 12.6\pm1$. This is much less than in the other data sets. However, Richardson \textit{et al.} conducted a study of 5700 patients hospitalized in NYC area and found even lower values: the overall length of stay was only 4.1 days (2.3-6.8)~\cite{Richardson-2020}. Possible causes of discrepancy are (i) over-simplification of the present approach or (ii) the limited interval of the study by Richardson {\it et al.}, between March 1 and April 4, 2020, shortly before the peak of the outbreak.

Italy, Switzerland, and Belgium display similar figures for $t_{ICU}$: slightly more than 16 days in Italy, slightly less in Belgium, and between 13 and 17.25 for Switzerland. In France, again, $t_{ICU}$ appears to be significantly longer.

\section{Conclusion}

In this paper, we have shown how the simplest of all epidemiological models suffices to match macroscopic data with almost perfection.
Having not let the pandemic evolve freely, political decision makers have curbed the outbreaks in a way that can be modelled by simple analytical formulas. These provide mathematical models for bed occupation numbers as a function of time that can be fitted very closely to the data supplied by health agencies. The fitting procedure yields estimates of some important epidemiological parameters of COVID-19. Assuming values of the decay rate $\Gamma$ of the outbreak, one derives estimates of the time from contamination to hospital, the time in hospital, and the time in ICU. Numbers obtained are consistent with previously published values. Ranges of confidence are given, but they are conditioned by  the value $\Gamma$. Still, interesting trends are observed, notably the much shorter hospitalization time inferred for NYC compared to the other geographical areas. Overall, a great disparity of values is observed depending on geographical location. Local circumstances, in the form of numbers of available beds, massive public gathering, peculiarities in the lockdown measure,  and also public awareness, certainly have impacted the parameters of the local outbreaks.  This should be taken into account in epidemiological models (see, e.g.~\cite{Chowdhury-2020}). If only macroscopic data, such as those analyzed here, are available, it appears unnecessary to resort to more complicated models than SIR or close variants thereof.

\subsubsection*{Acknowledgement}
G.K. is a Research Associate with the Belgian Fonds de la Recherche Scientifique (FNRS). Thanks to Leonardo Lanari and Daniela Iacoviello for  helping to  get  Italian data. \\


\end{document}